\documentclass[conference]{IEEEtran}
\usepackage{graphicx} 
\usepackage{amsmath}
\usepackage{braket}
\usepackage[english]{babel}
\usepackage{adjustbox}
\usepackage{hyperref}
\usepackage{cleveref}
\usepackage{tikz}
\usetikzlibrary{quantikz}
\usepackage{amsfonts}

\title{Agnostic Dynamical Decoupling for Single-Qubit Gates}
\author{\IEEEauthorblockN{Gumaro Rendon}
\IEEEauthorblockA{\textit{Fujitsu Research of America, Inc}\\
Santa Clara, US \\
gumaro.rendon@gmail.com}
}
\date{\today}

\begin{document}

\maketitle

\begin{abstract}
    We introduce a method for designing smooth single-qubit control pulses that implement a desired gate while suppressing the effect of unknown static error sources to first order. Unlike dynamically corrected gate constructions that require prior knowledge of the noise model, the present approach is agnostic to the detailed form of the target-bath interaction. The method parametrizes the control propagator through an auxiliary matrix expansion over orthogonal basis functions and enforces decoupling through algebraic orthogonality and equal-norm constraints on the expansion coefficients. These conditions guarantee that the leading Magnus contribution of an arbitrary static interaction reduces to a term proportional to the identity on the target system, thereby cancelling first-order error effects independently of the microscopic origin of the noise. We further show that the same construction suppresses, to first order, mediated couplings between simultaneously controlled qubits when their interaction occurs through intermediate environmental degrees of freedom, yielding effective second-order decoupling of the induced inter-qubit interaction. By using a discrete cosine transform parametrization, the pulse-synthesis problem is cast into a numerically stable constrained optimization with a minimal number of free parameters. Numerical examples for $R_z$ rotations and random single-qubit unitaries demonstrate smooth control fields that realize the target gates while remaining robust against arbitrary static single-qubit noise and mediated multi-qubit couplings. These results provide a hardware-friendly route toward noise-agnostic dynamically corrected single-qubit gates.
\end{abstract}

\section{Introduction}

Traditional error correction in quantum computing relies on encoding logical qubits on a larger set of physical ones from which they extract the occurred errors through syndrome measurement. These methods, which we call error correcting codes, need a certain upper threshold on the error rates for them to overcome the combinatorial factor from the different possibilities for errors; that threshold fulfilled, errors decrease exponentially with the number of quantum resources by concatenating recursive error correcting codes \cite{nielsen2010quantum}. This paradigm of error correction is best seen as a reactive error correction paradigm, because one performs measurements first to characterize the error that has occurred.

The strategy here is different, or complimentary. The pulses to control the qubits is engineered in such a way that errors are automatically cancelled at the end of the pulse. The approach was first thought of in the context of Nuclear Magnetic Resonance (NMR)~\cite{1950PhRv...80..580H,1954PhRv...94..630C,1958RScI...29..688M,haeberlen1976high,2004RvMP...76.1037V}. This was for extending the coherence time of qubits in stand-by. In~\cite{LEVITT198661,1989JMagR..81..423G,1994JMagR.109..221W,2003PhRvA..67d2308C,2009Natur.458..996B,2010PhRvL.104i0501K,JONES201191,2012Natur.484...82V,2012NatCo...3..997W,2013NJPh...15i5004G,2013PhRvL.110n0502K,2017PhRvL.118o0502C}, this idea is extended for rotations to qubit states while cancelling out errors. For an extensive review on dynamically corrected gates in general see ~\cite{Barnes_2022}.

New insight has also come out of novel frameworks like Space Curve Quantum Control (SCQC). The qubit controls are cast to space curves, where for example, the cancellation of first-order errors come out as closed-loop condition on the curves~\cite{Zeng_2018}. 

However, for this last framework one must assume the form of the static error term, which entails characterization of the hardware error sources. The precision to which we know the form of the noise is going to impact how well we cancel it.

There is a new work~\cite{PhysRevLett.132.193801} that aimed at canceling the effects of a general noise term up to first order. They successfully accomplish this, but their pulse design may require discontinuous pulses that may be hard to realize in real hardware. 

Here, we introduce a new method which is also agnostic to the source of the error and for which we cancel its effects to first-order in an active manner. Moreover, we demonstrate the method we provide cancels, to first-order, the entangling interactions between the environment and the target system as well as entangling interactions among the qubits if they interact through intermediary degrees of freedom, like photons mediate interaction between electrons. Finally, the method presented here uses smooth pulses which are more amenable for real hardware implementations.

\section{Method}

Suppose we have a Hamiltonian for a target system and a bath of the form
\begin{align}
   H = H_{I} + H_{0}.
\end{align}
Here, we split the general Hamiltonian in two parts:
\begin{align}
    H_{0} = H_{C} (t),
\end{align}
\begin{align}
    H_{I} = H_{T} + H_{TB} + H_{B}.
\end{align}
We can assume a sufficiently large target-bath system such that the Hamiltonian apart from the control is time-independent/autonomous. In the Heisenberg-picture, the interaction term is defined through:
\begin{align}
    \mathcal{H}_{I}(t) = U_C^\dagger(t) H_{I} U_C(t),
\end{align}
where
\begin{align}
    U_C = \exp_{\mathcal{T}}{\mathrm{i}\int H_{C}(t) \mathrm{d} t }.
\end{align}
The leading-order term in the Magnus expansion of this Heisenberg-picture Hamiltonian is:
\begin{align}
    A_1(t) = \int^{t}_0 \mathcal{H}_I(t_1) \mathrm{d}t_1.
\end{align}
In order to parametrize/construct our unitary matrix $U_C$ we will first define the auxiliary matrix
\begin{align}
    V_C(\lambda) = \rho_0 (\lambda) I_K + \sum_{k=1} \rho_k (\lambda) \Lambda_k,  
\end{align}
where $\lambda \in [0,1]$, and the generators $\Lambda_k$ are normalized such that
\begin{align}
    \mathrm{Tr}\left(\Lambda_j \Lambda_k\right) = K\delta_{jk}.
\end{align}
We will construct $U_C$ through $V_C$ the following way
\begin{align}
    U_C = g(\lambda) V_C(\lambda).
\end{align}

Unitarity requires:
\begin{align}
    U_C^\dagger U_C = I.
\end{align}
For $U(2)$, this constraint is fulfilled with

\begin{align}
    g(\lambda) = \frac{1}{\| \rho \|_s},
\end{align}
where
\begin{align}
    \| \rho \|_s = {}_s\sqrt{\sum_\kappa \rho_\kappa^\star \rho_\kappa },
\end{align}
and 
\begin{align}
    \rho_0 &= e^{i\phi}f_0 \cr
    \rho_k &= i e^{i\phi}f_k,
\end{align}
where $ \phi = \arg \rho_0 $ and $f\in \mathbb{R}$.  Here the subscript $s$ denotes that for the square root we will choose the positive solution when $\lambda=0$ and if the argument, $\sum_\kappa \rho_\kappa^\star \rho_\kappa$, becomes zero as we increase $\lambda$, we choose the solution branch that leaves $\| \rho \|_s$ analytic.

With these choices, $U_C$ and $V_C$ become
\begin{align}
    V_C(\lambda) = e^{i \phi}\left(f_0 (\lambda) I_K + i\sum_{k=1} f_k (\lambda) \Lambda_k  \right)
\end{align}
\begin{align}
    U_C(\lambda) = \frac{V}{\left\|f\right\|_s}
\end{align}
\begin{align}
    \left\|f\right\|_s = {}_s\sqrt{\sum_\kappa f_\kappa f_\kappa }
\end{align}
The relation between time, $t\in [0,T]$ and the parametrization variable $\lambda$ is:
\begin{align}
    \frac{\mathrm{d}t}{\mathrm{d}\lambda} = \frac{T}{\int_0^1 \left\|f\right\|_s^2 \mathrm{d} \lambda }\left\|f\right\|_s^2.
\end{align}
The variable $\phi$ would have no observable effect, thus we choose to set $\phi=0$ such that $U_C$ belongs to the group $SU(2)$.

We can enforce the boundary conditions on $U_C$ also on $V_C$ and automatically $\|f(\lambda)\|_s\Big\vert_{\lambda=0,1}=1$ . That is, at initial time $\lambda=0$ ($t=0$), we have
\begin{align}
    V_C(\lambda)\Big\vert_{\lambda=0} = I_K,
\end{align}
which means $f_0(\lambda)\Big\vert_{\lambda=0} = 1$ and $f_k(\lambda)\Big\vert_{\lambda=0} = 0$. We can also enforce the conditions at the final time through
\begin{align}\label{eq:boundary_constraints}
    f_0(\lambda)\Big\vert_{\lambda=1} = \frac{1}{K}\mathrm{Tr} \left(U\right), \cr
    f_k(\lambda)\Big\vert_{\lambda=1} = \frac{1}{i K}\mathrm{Tr} \left(\Lambda_k U \right).
\end{align}
Now, we would like to cancel out first-order errors, so we need to impose the following condition: Given two initial and final states on the bath system, the interaction term must be proportional to the identity.

\begin{align}
   A_{1,\psi_1,\psi_2} &= \bra{\psi_2} A_1 \ket{\psi_1}  \cr 
   &= \int^T_0 U_C^\dagger (\lambda) \langle \psi_1 \vert H_I  \vert \psi_2  \rangle U_C(\lambda) \mathrm{d}t \propto I. \cr 
\end{align}
After changing the variables from $\lambda \to T$, $A_1$ becomes:
\begin{align}
    A_1 &= \int^T_0 U_C^\dagger (\lambda) H_I U_C(\lambda)\mathrm{d}t \cr
    &= \int^1_0 \frac{1}{\left\|f\right\|_s^2} V_C^\dagger (\lambda) H_I V_C(\lambda) \frac{\mathrm{d}t}{\mathrm{d}\lambda} \mathrm{d}\lambda \cr 
    &= \frac{T}{\int_0^1 \left\|f\right\|_s^2 \mathrm{d} \lambda }\int^1_0  V_C^\dagger (\lambda) H_I V_C(\lambda) \mathrm{d}\lambda.
\end{align}
%
Thus, $A_{1,\psi_1,\psi_2}$ can be written the following way:
\begin{align}
    A_{1,\psi_1,\psi_2} &= \bra{\psi_2} A_1 \ket{\psi_1} \cr 
    &= \frac{T}{\int_0^1 \left\|f\right\|_s^2 \mathrm{d} \lambda }\int^1_0 V_C^\dagger (\lambda) \bra{\psi_2} H_I \ket{\psi_1} V_C(\lambda) \mathrm{d}\lambda \cr
    &= \frac{T}{\int_0^1 \left\|f\right\|_s^2 \mathrm{d} \lambda }\bigg( \phi_0 I \cr 
    &\quad + \int^1_0 V_C^\dagger (\lambda) \left(\sum_k \phi_k \Lambda_k \right) V_C(\lambda) \mathrm{d}\lambda  \bigg) \cr 
\end{align}
Now, we insert the expressions for $V_C$ in terms of $f_\kappa$'s and the generators:
\begin{align}
A_{1,\psi_1,\psi_2}&=  \frac{T}{\int_0^1 \left\|f\right\|_s^2 \mathrm{d} \lambda } \Bigg(\phi_0 I \cr
&\qquad+ \int^1_0  \left(f_0 (\lambda) I_K- i\sum_{k_1=1} f_{k_1} (\lambda) \Lambda_{k_1}  \right) \cr
&\qquad\qquad\times\left(\sum_{k_2} \phi_{k_2} \Lambda_{k_2} \right) \cr
&\qquad\qquad\times\left(f_0 (\lambda) I_K+i \sum_{k_3=1} f_{k_3} (\lambda) \Lambda_{k_3}  \right) \mathrm{d}\lambda \Bigg). \cr  
\end{align}
Expanding:
\begin{align}
A_{1,\psi_1,\psi_2}&= \frac{T}{\int_0^1 \left\|f\right\|_s^2 \mathrm{d} \lambda }\Bigg(\phi_0 I +  \left(\sum_{k_2} \phi_{k_2} \Lambda_k \right)\int^1_0 f_0^2 \mathrm{d}\lambda \cr
&\quad+ i\left(\sum_{k_2} \phi_{k_2} \Lambda_{k_2} \right)\int^1_0 f_0  \left(\sum_{k_3} f_{k_3} \Lambda_{k_3} \right)  \mathrm{d}\lambda \cr 
&\quad-i\int^1_0 f_0  \left(\sum_{k_1} f_{k_1} \Lambda_{k_1} \right)  \mathrm{d}\lambda \left(\sum_{k_2} \phi_{k_2} \Lambda_{k_2} \right) \cr 
&\quad +   \int^1_0  \left(\sum_{k_1=1} f_{k_1} (\lambda) \Lambda_{k_1}  \right) \cr
&\quad\times\left(\sum_{k_2} \phi_{k_2} \Lambda_{k_2} \right)  \left(\sum_{k_3=1} f_{k_3} (\lambda) \Lambda_{k_3}  \right) \mathrm{d}\lambda\Bigg). \cr
\end{align}
We now assume an orthogonal polynomial expansion of the form:
\begin{align}
    f_\kappa(\lambda) = \sum^{M-1}_{j=0} a_{\kappa,j} p_j(\lambda),
\end{align}
such that
\begin{align}
    \int p_l (\lambda) p_m (\lambda) \mathrm{d} \lambda = c_l\delta_{lm}.
\end{align}
With this, 
\begin{align}
    A_{1,\psi_1,\psi_2} &=  \frac{T}{\int_0^1 \left\|f\right\|_s^2 \mathrm{d} \lambda } \Bigg(\phi_0 I +  \left(\sum_{k_2} \phi_{k_2} \Lambda_k \right) \mathbf{a}_0 \odot \mathbf{a}_0 \cr
    &+ i\left(\sum_{k_2} \phi_{k_2} \Lambda_{k_2} \right)  \left(\sum_{k_3}  \mathbf{a}_0 \odot \mathbf{a}_{k_3}  \Lambda_{k_3} \right)   \cr 
&\quad-i \left(\sum_{k_1} \mathbf{a}_0 \odot  \mathbf{a}_{k_1} \Lambda_{k_1}  \right)   \left(\sum_{k_2} \phi_{k_2} \Lambda_{k_2} \right) \cr 
&\qquad +   \sum_{k_1=1}\sum_{k_2}\sum_{k_3=1}\Lambda_{k_1}\Lambda_{k_2}\Lambda_{k_3} \phi_{k_2}   \mathbf{a}_{k_1} \odot \mathbf{a}_{k_3}\Bigg),\cr 
\end{align}
where
\begin{align}
    \mathbf{a}_i\odot \mathbf{a}_j = (\mathbf{a}_i)_k C_{kl} (\mathbf{a}_j)_l
\end{align}
and
\begin{align}
    C_{kl} = \delta_{kl}c_k.
\end{align}
We will establish the form of $C_{kl}$ once we choose the orthogonal polynomials. Now, we propose the orthogonality between the coefficients vectors. That is,
\begin{align}
    \mathbf{a}_{\kappa} \odot \mathbf{a}_{\sigma} = 0 \qquad \text{ for } \kappa \neq \sigma.
\end{align}
This means
\begin{align}
    A_{1,\psi_1,\psi_2} &=  \frac{T}{\int_0^1 \left\|f\right\|_s^2 \mathrm{d} \lambda } \Bigg(\phi_0 I + \left(\sum_{k_2} \phi_{k_2} \Lambda_k \right) \mathbf{a}_0 \odot \mathbf{a}_0 \cr
    &+ \sum_{k_1}\sum_{k_2}\Lambda_{k_1}\Lambda_{k_2}\Lambda_{k_1} \phi_{k_2}   \mathbf{a}_{k_1} \odot \mathbf{a}_{k_1} \Bigg)
\end{align}
Now, we assume
\begin{align}
    \mathbf{a}_{k_1} \odot \mathbf{a}_{k_1} = \mathbf{a}_{k_2} \odot \mathbf{a}_{k_2} \qquad\text{for }k_1,k_2\geq 1.
\end{align}
Thus,
\begin{align}
    A_{1,\psi_1,\psi_2} &= \frac{T}{\int_0^1 \left\|f\right\|_s^2 \mathrm{d} \lambda } \Bigg (\phi_0 I +  \left(\sum_{k_2} \phi_{k_2} \Lambda_k \right) \mathbf{a}_0 \odot \mathbf{a}_0 \cr
    &+  \mathbf{a}_{1} \odot \mathbf{a}_{1} \sum_{k_2}\phi_{k_2}\sum_{k_1}\Lambda_{k_1}\Lambda_{k_2}\Lambda_{k_1}   \Bigg)
\end{align}
For the last term, we can use
\begin{align}
    \sum_{k_1}\Lambda_{k_1}\Lambda_{k_2}\Lambda_{k_1}   = -  \Lambda_{k_2}
\end{align}
in order to obtain
\begin{align}
    A_{1,\psi_1,\psi_2} &=  \frac{T}{\int_0^1 \left\|f\right\|_s^2 \mathrm{d} \lambda } \Bigg(\phi_0 I  \cr
    &+\left(\sum_{k_2} \phi_{k_2} \Lambda_k \right) \left(\mathbf{a}_0 \odot \mathbf{a}_0 -  \mathbf{a}_{1} \odot \mathbf{a}_{1} \right)\Bigg).  
\end{align}
This means we also need
\begin{align}
    \mathbf{a}_{\kappa_1} \odot \mathbf{a}_{\kappa_1} = \mathbf{a}_{\kappa_2} \odot \mathbf{a}_{\kappa_2} \qquad\text{for }\kappa_1,\kappa_2\geq 0.
\end{align}

\subsection{Solution}

By choosing to sample or assign values to $f_i(\lambda)$ at $M$ chosen nodes $\lambda_j$, we establish the following linear system of equations:
\begin{align}
A\mathbf{a}_i=\mathbf{b}_i
\end{align}
such that the necessary boundary conditions for $f_i$ at $\lambda \in \{0,1\}$. Recall that,
\begin{align}
    f_i(\lambda) = \sum^{M-1}_{j=0} \left(\mathbf{a}_i\right)_j p_j (\lambda).
\end{align}
which we expect to interpolate through
\begin{align}\label{eq:fibi}
    f_i(\lambda_j) = \left(\mathbf{b}_i\right)_j
\end{align}
for a set of $M$ collocation points $\lambda_j$.

 With this the solution for $\mathbf{a}_i$ of the following form
\begin{align}
    \mathbf{a}_i = A^{-1} \mathbf{b}_i
\end{align}
where $A^{-1}$ is the inverse of $A$. Apart from the boundary constraints (\Cref{eq:boundary_constraints}), we need to enforce the following orthogonality relation as we discussed before:
$$
\mathbf{a}_i \odot \mathbf{a}_j = \mathbf{a}_i \odot \mathbf{a}_i \delta_{ij}.
$$
Thus, we will assume that $M$ is sufficiently large to accommodate these constraints as well. And the remaining values of $\mathbf{b}_i$ not fixed by the boundary conditions at $\lambda=0$ and $\lambda=1$ are variationally optimized to fulfill this relation.

One such set of polynomials is:
\begin{align}
    p_n(x) = \sqrt{d_n}\cos{\pi n x} / \sqrt{M-1}
\end{align}
where
\begin{align}
    d_n = \begin{cases}
        1 & n =0,M-1 \\
        2 & n \neq 0,M-1
    \end{cases}
\end{align}
With this, we have set
\begin{align}\label{eq:ck}
    c_n = \begin{cases}
        \frac{1}{2(M-1)} & n = M-1 \\
        \frac{1}{M-1} & n \neq M-1.
    \end{cases}
\end{align}

The matrix $A$ looks the following:
\begin{align}
    \left(A\right)_{ij} = \sqrt{d_j}\cos\left(\pi j \lambda_i\right) / \sqrt{M-1}
\end{align}
To ensure orthogonality of matrix we choose the set of nodes:
\begin{align}
    \lambda_i = \frac{i}{M-1} \text{ for }i=0,1,\dots,M-1. 
\end{align}
It will be however, more convenient to re-scale the first and last-rows of the system of equations which will make more evident the stability of the system:
\begin{align}\label{eq:linear_system}
    \widetilde{A} \mathbf{a}_i &= \widetilde{\mathbf{b}}_i \cr
    \left(A\right)_{ij} &= \sqrt{\frac{ e_i d_j}{ \left(M-1\right)}}\cos\left(\frac{\pi j i}{M-1}\right) \cr 
    \left(\widetilde{\mathbf{b}}_i\right)_j &= \sqrt{e_j}\left(\mathbf{b}_i\right)_j
\end{align}
where
\begin{align}
        e_j = \begin{cases}
        1/\sqrt{2} & j = 0,M-1 \\
        1 & j \neq 0,M-1.
    \end{cases}
\end{align}
After this, the matrix $\widetilde{A}$ is the orthonormal DCT-1 transform and as such,
\begin{align}\label{eq:orthonorm}
    \widetilde{A}^{-1} = \widetilde{A}^{\top}.
\end{align}

From \cref{eq:linear_system} and \Cref{eq:orthonorm} we have that
\begin{align}
    \widetilde{\mathbf{b}}_i \cdot \widetilde{\mathbf{b}}_j = \mathbf{a}_i \cdot \mathbf{a}_j
\end{align}
We recall that:
\begin{align}
   \mathbf{a}_i \odot \mathbf{a}_j &= \left(\mathbf{a}_i\right)_{k}  C_{kl} \left( \mathbf{a}_j \right)_{l} \cr 
   &= \left(\mathbf{a}_i\right)_{k} c_k \delta_{kl} \left( \mathbf{a}_j \right)_{l} \cr 
   &= \left(\mathbf{a}_i\right)_{k} \sqrt{c_k} \delta_{kl} \sqrt{c_l} \left( \mathbf{a}_j \right)_{l},
\end{align}
and we note from the distribution of the coefficients $c_k$ (See \Cref{eq:ck}), that only the case $k=M-1$ is the only different one. With this, if we impose the restriction that the last coefficients in $\mathbf{a}_i$ for each $i$ is zero, the orthogonality condition becomes:
\begin{align}\label{eq:new_orth}
   \mathbf{a}_i \odot \mathbf{a}_j = \frac{\mathbf{a}_i \cdot \mathbf{a}_j}{M-1}.
\end{align}
Here the condition that the latest coefficient in $\mathbf{a}_i$ for any $i$ is zero is imposed through
\begin{align}\label{a_last_zero}
    \left(\mathbf{a}_i\right)_{M-1} &=\sum_k\left(A^{\top}\right)_{(M-1),k} \left(\widetilde{\mathbf{b}}_i\right)_k = 0.
\end{align}
Thus, under this constraint:
\begin{align}
\widetilde{\mathbf{b}}_i \cdot \widetilde{\mathbf{b}}_j &= \mathbf{a}_i \cdot \mathbf{a}_j= (M-1)\mathbf{a}_i \odot \mathbf{a}_j 
\end{align}
The new orthogonality and norm conditions can be passed to the vectors $\widetilde{\mathbf{b}}_i$ through:
\begin{align}
\widetilde{\mathbf{b}}_i \cdot \widetilde{\mathbf{b}}_j &= \delta_{ij}\mathbf{a}_0 \cdot \mathbf{a}_0.
\end{align}
In order to parametrize these orthogonal vectors efficiently, we can embed these $K^2$ vectors, $\widetilde{\mathbf{b}}_i$, into the columns of an orthogonal matrix of dimensions $M \times M$
\begin{align}
    B,
\end{align}
such that the vectors $\widetilde{\textbf{b}}_i$ are embedded in the matrix $B$ the following way:
\begin{align}\label{eq:embedding}
    B = \begin{pmatrix}
        \widetilde{\textbf{b}}_0 & \widetilde{\textbf{b}}_1 & \cdots & \widetilde{\textbf{b}}_{K^2-1} \cdots 
    \end{pmatrix}
\end{align}
which in turn we can parametrize with:
\begin{align}
    B = o_0^2 e^{\sum^{M-1}_{i=0}\sum^{M-1}_{j = i+1} o_{ij} \Sigma^{(ij)}}
\end{align}
where $\Sigma^{(ij)}$ are skew-symmetric generators of the orthogonal group $O(M)$. We choose the following basis:
\begin{align}
    \left(\Sigma^{(ij)}\right)_{kl} = \begin{cases}
        1 &\quad\text{for } (k,l)=(i,j) \cr
        -1 &\quad\text{for }(k,l)=(j,i) \cr
        0 &\quad\text{otherwise}.
    \end{cases}
\end{align}
With this parametrization, we have that
\begin{align}
    B^{\top} B = o_0^2 I_M,
\end{align}
thus, $\mathbf{a}_0 \cdot \mathbf{a}_0 = o_0^2$, where $o_0$ is going to be a free parameter in our optimization procedure. The embedding in \Cref{eq:embedding} implies the constraints:

\begin{align}\label{eq:B_conditions}
    B_{0,j} &= f_j(0)/\sqrt{2},\cr
    B_{M-1,j} &= f_j(1)/\sqrt{2}.
\end{align}

\section{Obtaining Controls from $U_C(\lambda)$}

\begin{align}
    H = i\frac{\mathrm{d}}{\mathrm{d} t} \left(U_C (t)\right) U_C(t)^\dagger
\end{align}
\begin{align}
    g_k(t) = {\rm Tr}\left(H(t)\Lambda_k\right)/K
\end{align}
\begin{align}
    H(t) = \sum g_k(t) \Lambda_k
\end{align}

\begin{align}
    U_C(\lambda) = \frac{V_C}{\left\|f\right\|_s} =  \frac{1}{\left\|f\right\|_s}\left(f_0 (\lambda) I_K + i\sum_{k=1} f_k (\lambda) \Lambda_k \right),
\end{align}
\begin{align}
    \frac{\mathrm{d}}{\mathrm{d}\lambda}\frac{1}{\left\|f\right\|_s}= -\frac{1}{\left\|f\right\|_s^{3}} \sum_\kappa f_\kappa(\lambda) \dot{f}_\kappa (\lambda)
\end{align}
\begin{align}
    \frac{\mathrm{d}}{\mathrm{d}\lambda}U_C(\lambda) = \frac{\dot{V}_C(\lambda)}{\left\|f\right\|_s} -\frac{V_C}{\left\|f\right\|_s^{3}} \sum_\kappa f_\kappa(\lambda) \dot{f}_\kappa (\lambda)
\end{align}

\begin{align}
    \frac{\mathrm{d}t}{\mathrm{d}\lambda} = \frac{T}{\int_0^1 \left\|f\right\|_s^2 \mathrm{d} \lambda }\left\|f\right\|_s^2
\end{align}

We also enforce
\begin{align}
 g_k(1) = 0 
\end{align}
which are $4$ constraints in total. The constraints $g_k(0)=0$ are automatically fulfilled by the definition of $U_C$. That, combined with the other $6$ from \Cref{eq:B_conditions} and the $4$ makes $14$. The number of generators for $M=6$ is $6*5/2=15$ plus one from the norm parameter, $o_0$. Using $M=5$ would not give enough free parameters to enforce all the constraints, so $M=6$ is the absolute minimum. 

\subsection{Control Factorization}

It is clear that the time dependence of the control $U_C$ on the target system has no effect on the bath term, $H_B$, even if we promote it to a time dependent term. Thus, we can simultaneously control different systems without affecting each other by just targeting other sectors of the bath which might be the other parts of the quantum register.

The problem comes when the magnitude of the controls is large, it impedes the convergence of the Magnus expansion

\begin{align}
    \int^T_0 \| H_C \| {\rm d} s \leq \pi
\end{align}
However, if the interaction between system qubits/qudits is mediated by another bath degree of freedom, we achieve the quadratic error correction as well. This is expected to be the case through most hardware realizations, for example, electron-spin only interact with each other through photons.

\section{Results and Simulations}

\begin{figure}[!ht]
    \centering
    \includegraphics[width=0.75\linewidth]{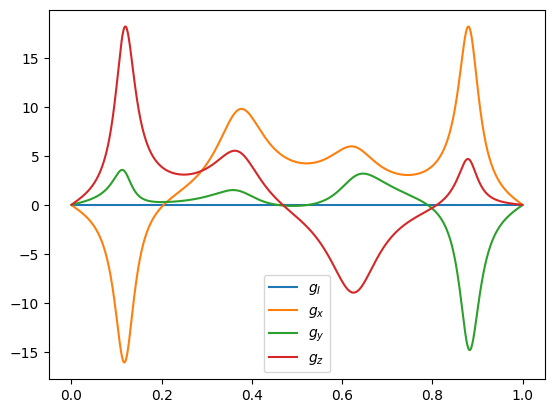}
    \caption{The three smooth pulses needed to control all the generators, $\sigma_x$, $\sigma_y$, and $\sigma_y$ of SU(2). The gate implemented is an $R_z(\theta)$ gate with $\theta = 0.025 \pi$.}
    \label{fig:pulses}
\end{figure}

We start off by focusing on the controls of a target 1-qubit system. Like we have discussed, the controls obtained won't depend on the error source term. For a $R_z$ gate we obtain the pulses shown in \Cref{fig:pulses}.

\begin{figure}[!ht]
    \centering
    \includegraphics[width=0.75\linewidth]{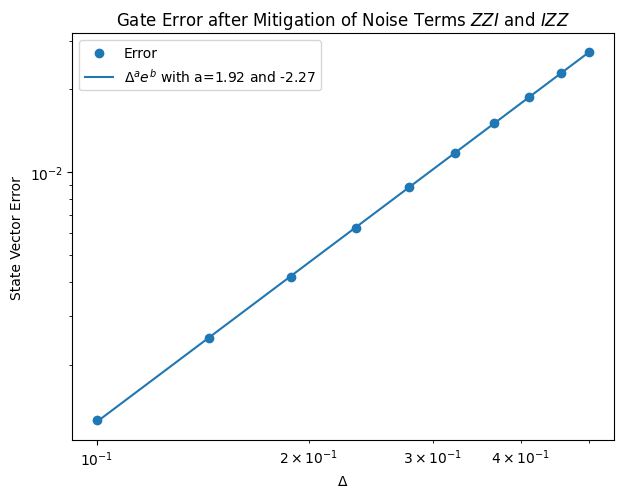}
    \caption{Simultaneous control of 2 qubits while in presence of noise terms that couple them to an intermediate qubit. This, to demonstrate the second-order decoupling of said target qubits from one another for any intermediary quantum degrees of freedom. The gate implemented is an $R_z(\theta)$ gate with $\theta = 0.025 \pi$.}
    \label{fig:simult_error}
\end{figure}

\begin{figure*}[t]
    \centering
    \begin{quantikz}[row sep=0.7cm, column sep=0.9cm]
    \lstick{$q_1$} & \gate{e^{i \Delta t H_C(t_1)}} & \gate[2]{e^{i\delta \Delta t ZZ}} & \qw   & \gate{e^{i \Delta t H_C(t_2)}} & \qw  & \dots   &  \\
    \lstick{$q_m$} & \qw & \qw  & \gate[2]{e^{i \delta \Delta t  ZZ}}  & \qw &  \qw    & \dots &  \\
    \lstick{$q_2$} & \gate{e^{i \Delta t H_C(t_1)}}& \qw      & \qw  & \gate{e^{i \Delta t H_C(t_2)}} & \qw  & \dots  &
    \end{quantikz}

    \caption{Trotterized Representation of the simultaneous control of two-qubits. The target qubits for the $R_z$ gates are $q_0$ and $q_2$ and $q_1$ represents the bath qubit. The choice of error terms are $IZZ$ and $ZZI$}
    \label{fig:schematic_trotter}
\end{figure*}
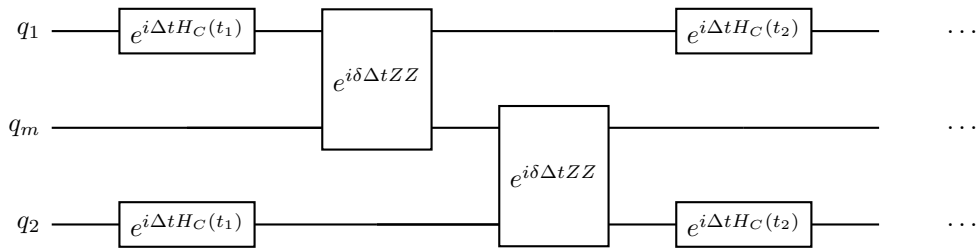

With these same pulses we can control simultaneously two qubits to perform an $R_z$ gate on each while still decoupling them up to second order. One can see \Cref{fig:simult_error} the effects from the noise corrected to second-order. The noise terms are $ZZI$ and $IZZ$, which couple the target qubits with a one-qubit bath in the middle. See \Cref{fig:schematic_trotter} for a Trotterized representation of the simultaneous control of two-qubits in the presence of noise terms $ZZI$ and $IZZ$.

\begin{figure}[!ht]
    \centering
    \includegraphics[width=0.75\linewidth]{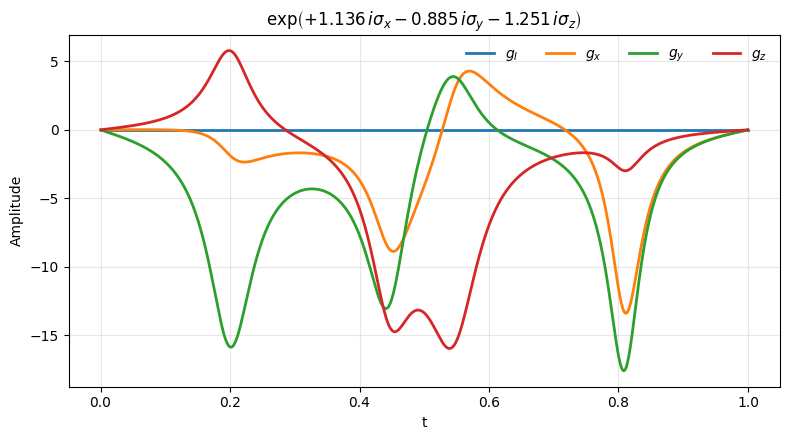}
    \caption{Smooth pulses generated when the gate implemented is chosen at random.}
    \label{fig:pulses_random_gate}
\end{figure}

\begin{figure}[!ht]
    \centering
    \includegraphics[width=0.75\linewidth]{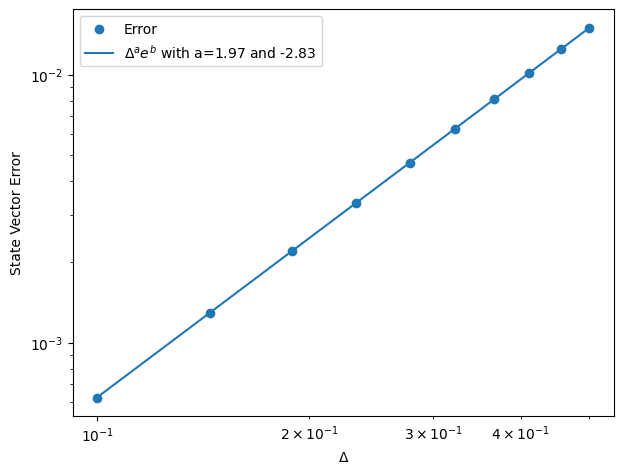}
    \caption{The decoupling also succeeds when the implemented gate (See \Cref{fig:pulses_random_gate}), and we obtain a second-order error.}
    \label{fig:error_random_gate_and_error}
\end{figure}

Moreover, we test the method with a random target 1-qubit gate along with a static random 1-qubit noise term \Cref{fig:pulses_random_gate}. With this, we demonstrate that the pulse design is completely agnostic to the form of the 1-qubit noise term \Cref{fig:error_random_gate_and_error}.

\section{Conclusion}

A method was presented for pulse design which decouples single qubits from their environment up to second order. This is done without knowledge of the form of the static interaction terms with the envirnoment and with pulses that are smooth and more amenable for real hardware implementations.
The work presented here only works for 1-qubit gates, extending this work to 2-qubit gates would be an important next step for the continuation of this work.

\section{Acknowledgements}

I would like to acknowledge the financial support of Error Corp throughout the development of this research project. I would also like to acknowledge the individual support of its CEO Dennis Lucarelli, who showed me the ropes of pulse engineering and dynamical decoupling and was always accommodating to my interests. 

\bibliographystyle{unsrt}
\bibliography{bibliography}

\end{document}